# Effects of using carbon Nano-tubes on thermal and ductility properties of bitumen

M. Faramarzi[1], M. Arabani[2], A. K. Haghi[3], V. Motaghitalab[4]

1. M. Sc. student, Department of civil engineering, University of Guilan, I. R. Iran
2. Professor, Department of civil engineering, University of Guilan, I. R. Iran
3. Professor, Department of textile engineering, University of Guilan, I. R. Iran
4. Assistant professor, Department of textile engineering, University of Guilan, I. R. Iran

Masoud.faramarzi67@gmail.com
Arabani@guilan.ac.ir
Haghi@guilan.ac.ir
motaghitalab@guilan.ac.ir

**Abstract**

New plans should be used to improve quality and to increase productivity and durability of conventional pavements. In this investigation, it has been attempted to promote technical characteristics of bitumen using carbon nanotubes as an additive. Wet and dry process methods are most practical ways of mixing CNF in AC. It was decided that the best method to adopt for this investigation was the dry process. In this study thermal and ductility properties of modified bitumen by 0.1, 0.5, and 1% carbon nano-tube content in bitumen were evaluated considering bitumen penetration, softening point, and ductility tests, then the results were compared to those of unmodified bitumen. It was found that adding carbon nano-tubes effects on thermal properties of bitumen by increasing the softening point and decreasing the bitumen penetration. It was also shown that bitumen ductility decreases by carbon nano-tubes modification process.

# 1. INTRODUCTION

Nanotechnology is a relatively new field in science dealing with structures that are on the nano-scale. To illustrate how miniscule the nano-scale is, the following comparison could be made: if a human hair has a diameter of a football field, a nano-sized particle would have the diameter of a pencil. In 1985, Kroto and Smalley first discovered buckminsterfullerene and since then this technology has evolved rapidly [1]. Nanosized particles have been used in numerous applications to improve various properties. However, due to many reasons including the cost of production and purification, nano-sized particles have seen very little use in the construction field. One of the promising additives in the construction field is the use of carbon nanotubes. In 1991, these materials were first characterized in depth by Iijima [2]. In general, carbon nanotubes are made of sheets of graphite that have rolled up to form a tubular structure. This is accomplished through various methods with a majority of them using electricity and an inert gas in an enclosed chamber [3]. There are two general types of carbon nanotubes: single-walled nanotubes (SWNTs) and multi-walled nanotubes (MWNTs). Because of the physics involved, MWNTs are easier and cheaper to produce [4]. However, they lack the strength found in SWNTs [4]. Nevertheless, SWNTs are found in bundles whereas MWNTs are found as individual molecules, allowing MWNTs to be dispersed efficiently in a material [4]. MWNTs are also stiffer than SWNTs and have a longer length, sometimes approaching the centimeter range [4]. The type and production of nanotubes will determine the diameter the materials which could range from 0.4nm (the smallest possible value) [5] up to several hundred nm [4]. This combination of long length and small diameter can lead to aspect ratios approaching 1,000,000:1. This gives rise to unique engineering characteristics such as some of these materials having a Young's Modulus of anywhere from 18 GPa to 68 GPa with a failure strain of 0.12 (12%) [6]. In addition, they can possess a tensile strength anywhere from 1.4 GPa to 2.9 GPa [6]. All of these properties make carbon nanotubes ideal candidates for improving the mechanical properties of various construction materials. Wu et al. [7] found that the modification applying nano-clay can not only improve the properties of bitumen but also lower costs to a great extent. In addition, Liu et al. [8] indicated that the montmorillonite nano-clay has been used successfully to strongly improve properties of polymer-modified binder. Even though and Xiao et al. [9, 10] considered that carbon nano-particle is beneficial in improving the rheological properties of modified asphalt binder after short-term and long-term aging procedures, there are still many challenges and difficulties to utilize these carbon nano-particles to enhance the properties of construction materials. There is a need for more research in utilizing nanotechnology to improve the rheological and engineering characteristics of unaged asphalt binders and develop some guidelines for their use in construction materials in general. The interaction of unaged modified binders is not well understood from the standpoint of binder properties and field performance. In addition, the research of nano particles in asphalt binders can expand the use of modified binders. Because of the complicated relationships of asphalt binders and carbon nano particles in the modified binders, detailed information will be beneficial to help obtain an optimum balance in the use of these materials. In the study presented in this paper process of making bitumen modified by carbon nano-tubes was explained. Penetration degree, softening point and ductility which are three classic tests on bitumen were done to evaluate modified bitumen rheological properties. Based on the obtained experimental results, the effects of CNTs on material properties were analyzed and discussed.

## 2. MATERIALS AND METHODS
### 2.1. Material

Materials used in the experimental investigation included a neat bitumen 60/70-penetration grade from Tehran mineral oil refinery with the physical properties listed in Table 1.

Table 1. Properties of used bitumen

| Purity Grade | Deflagration | Softening Point | Penetration Grade | Density |
|---|---|---|---|---|
| % | ºC | ºC | mm/10 | @ 25 ºC |
| 99 | 262 | 49 | 53 | 1.03 |

A commercially available multiwall CNTs with purity up to 95%, was used in this research. The image of Carbon nanotubes was presented in figure 1. Also the Characterization of Carbon nanotubes (CNTs) is showed at the Table 3.

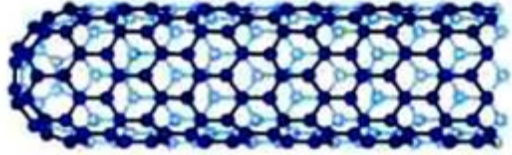

Fig.1: Carbon nanotubes (CNTs) image

Table.2 Carbon nanotubes (CNTs) Properties

| Properties | Unit | Value | Method of Measurement |
|---|---|---|---|
| Average Length | µm | 10-30 | TEM |
| Carbon Purity | % | 95 | TGA |
| Amorphous Carbon | - | * | HRTEM |
| Density | g/cm | 2.1 | - |

## 2.2. Experimental Procedure

Three different percentages of CNT were chosen to produce bitumen-CNT blends (0.1%, 0.5% and 1.0% by weight of the base binder). A simple shear mixing technique was employed to incorporate CNTs into the base bitumen not only because it is very convenient in laboratory operations, but also because it has the potential of being easily transferred to the industrial scale in hot mix asphalt plants. Following preliminary attempts in which different mixing times and temperatures were considered. The final mixing protocol adopted in the study consisted of two subsequent phases. the first phase in which CNTs were added and manually blended to the bitumen, and the second phase in which the bitumen-CNT blends were mixed with a mechanical stirrer operating at a speed of 1,550 rpm for a total time of 40 minutes in order to obtain satisfactory homogeneity. Scanning electron microscope (SEM) image of Carbon nanotubes (CNTs) are shown in figure 2. Also, the image of mechanical stirrer is presented in figure 3. During both phases of mixing, temperature was set at 160°C and kept constant by means of an oil bath, which was heated by a hot plate.

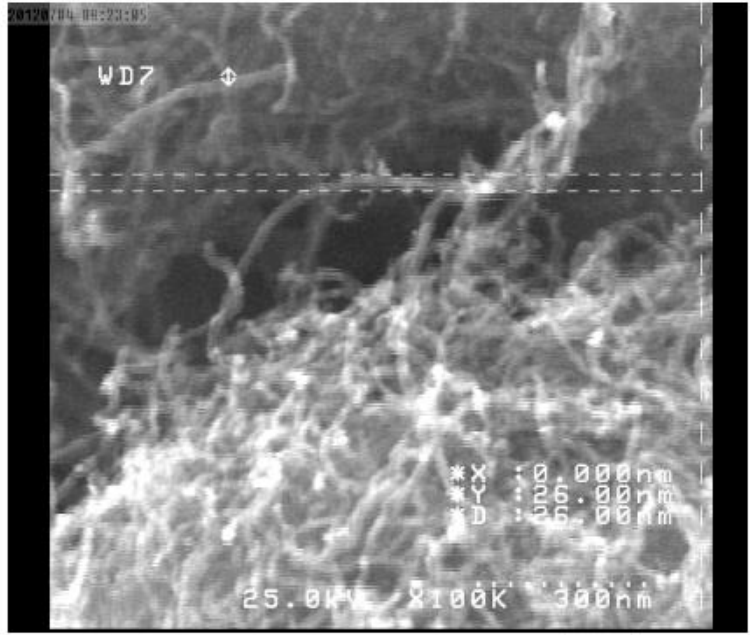

Fig. 2: Scanning electron microscope (SEM)

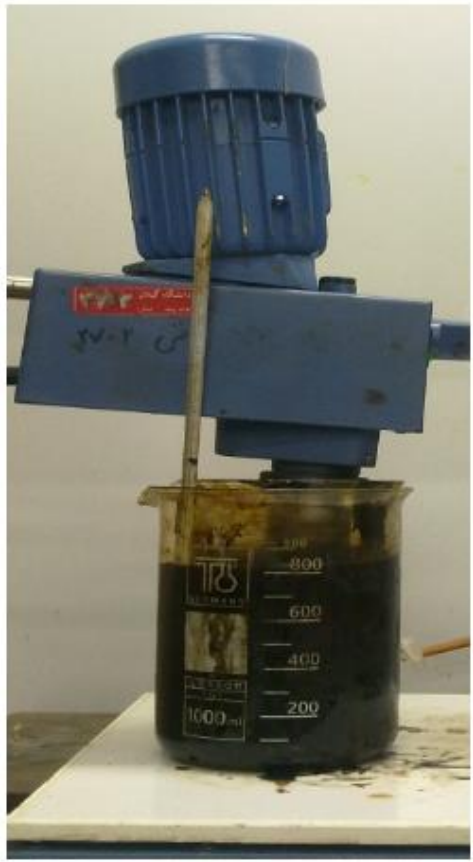

Fig.3: Mechanical stirrer used in this research image of Carbon nanotubes (CNTs)

### 2.3. Laboratory Tests

**Rheological Tests on Bitumen**

To determine the optimum content of carbon nanotubes (CNTs), empirical rheological tests carried out on conventional and modified bitumen with different CNTs content. In this study, the penetration degree test was done for Control sample and samples containing different percentages of carbon nanotubes. The empirical tests were performed according to the standard test procedures. The penetration test is an empirical test which measures the consistency (hardness) of asphalt at a specified test condition according to ASTMD5 standard. Also, for determination the softening point of bitumen, the ASTM-D36 was used. Ductility test according to the Standard Methods (IP. 32/55) was done and the sample at 25°C and with a speed of 5 cm/min is drawn.

## 3. Results and Analysis

### 3.1. Penetration Degree Test

Figure 5 shows a graph of the penetration degree changes. According to Figure 5, at first, it can be seen that there is no change in graph between control sample and 0.1% modified bitumen. It is because of the negative effect of aging, which happens during mixing process in high temperatures. With the addition of carbon nanotubes to bitumen, penetration degree of bitumen reduces. This reduction is due to the high surface density and high stability and tensile stability of carbon nanotubes. It is obvious that with increasing the amount of nanotubes, penetration degree will further be reduced. Finally, the produced bitumen can be used in warmer climates or areas with more traffic.

### 3.2. Softening Point Test

Figure 6 shows a graph of the softening point changes. As the result of aging during the mixing process and its effect on the bitumen thermal sensitivity a decrease is seen in the softening point at low levels of carbon nanotubes usage, but the higher levels of carbon nanotubes resulted in an increase, because the Young's modulus and high stability carbon nanotube causes the bitumen to show more stability against the flowing and thus increases the softening point. This bitumen can be used in areas with high average annual temperature or areas with more and heavier traffic.

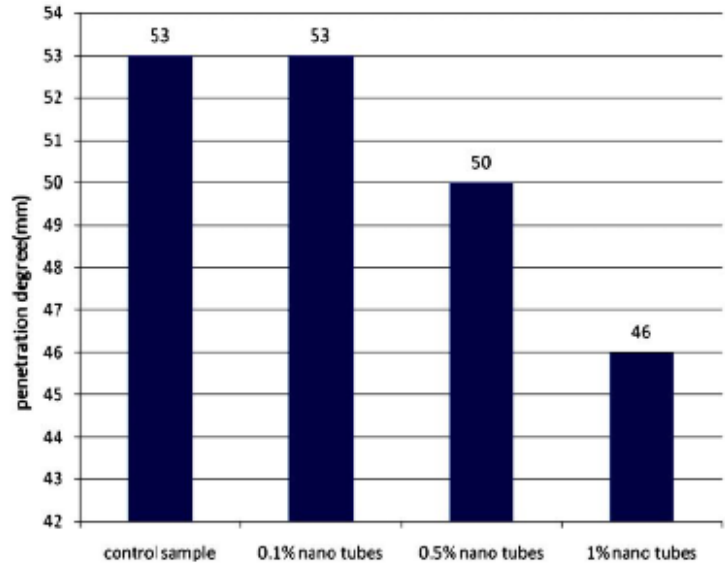

Fig.4: Comparison of penetration degree test for different samples

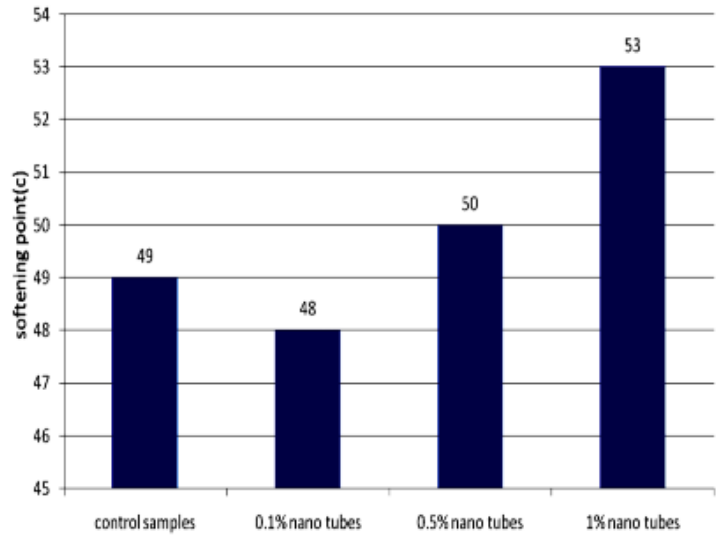

Fig.5: Comparison of softening point test results for different samples

### 3.3. Ductility Test

Figure 6 shows a graph of the ductility changes. In this case, with increasing carbon nanotubes, ductility properties will decrease much more. This behavior may be the result of chemical reaction and change in chemical structure as also pointed by Chile [24]. Checking by the ASTM-D113 [11], ductility of modified-samples by different percentages of CNT, was higher than the minimum of 50 cm. Bitumen which has acceptable ductility is more resistant to thermal cracking in low temperatures and shows good adherence so, could perfectly coat aggregates.

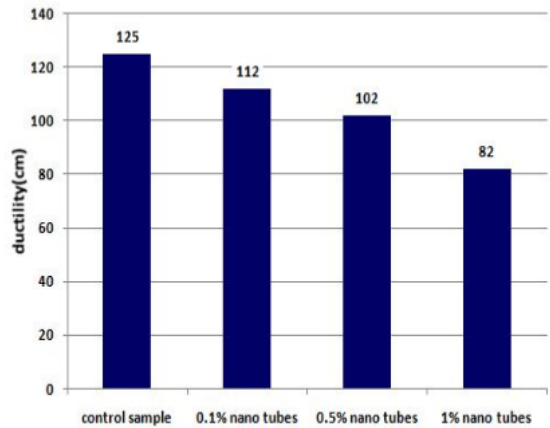

Fig.6: Comparison of ductility test results for different samples

### 4. CONCLUSIONS

Due to the increasing development of nanotechnology and special features of carbon nanotubes, you can use them as the ideal choice in asphalt mixtures. The aim of this study was experimental investigation on the effects of using carbon nano-tubes on rheological properties of bitumen binder. Based on the laboratory test results, the following conclusions were obtained:

Two samples, respectively, had the best results:

"A sample is containing 0.0005 carbon nanotubes by weight of bitumen".

"A sample is containing 0.001 carbon nanotubes by weight of bitumen".

The initial cost of both samples is higher than the control sample but for total cost, the amount and type of work should be investigated. When using modified bitumen in asphalt, due to its developed rheological properties, the asphalt layer thickness will be less than the usual and then the amount of total costs will decrease. A bitumen with minimal mechanical changes in its characteristics obtained that in comparison with conventional bitumen, not only in terms of physical properties is not much different, but also improves the quality of its chemical properties.